\documentclass[twoside]{dis09}
\usepackage[latin1]{inputenc}
\usepackage[dvips]{graphicx,epsfig,color}
\usepackage{wrapfig,rotating}
\usepackage{amssymb,amsmath,array}
\def\beq{\begin{equation}}
\def\eeq{\end{equation}}

\def\phid{\phi^{\dagger}}

\begin{document}
\title{PT Symmetry and Renormalization in Pomeron Model}

\author{Gian Paolo Vacca$^1$ 
%
\thanks{E-mail: vacca@bo.infn.it.}
%
\vspace{.3cm}\\
%
INFN sezione di Bologna \\
Via Irnerio 46, 40126 Bologna, Italy
}

\maketitle

\begin{abstract}
A novel perturbative analysis for the $2+1$ local supercritical field theory
of pomerons is developed. It is based on the PT symmetry of the model which
allows to study a similar Hamiltonian with the same real perturbative spectrum.
In the lowest non trivial order of perturbation theory
the pomeron interactions are shown to lead to the renormalization of the
slope. The appearance of a non local interaction for two pomeron states
is such that at small coupling only scattering states are present and
the spectrum of two particle states is not affected. 
\end{abstract}
\section{Introduction}

The high energy behavior of strong interaction in the Regge limit has been being studied
since more than forty years, at the beginning in the so called S-matrix theory approach
and subsequently using field theory models for the object describing the leading
behavior of the cross section, the Pomeron.
After Quantum Chromodynamics has been found to describe perturbatively many
aspects of strong interactions, the Regge limit of this theory has been
investigated leading to the so called BFKL physics~\cite{BFKL} and in general small $x$
physics, mainly analyzed in deep inelastic scattering experiments, as  the one
at HERA.

Theoretically in small $x$ QCD one can observe the emerging of an effective theory, which in
its more simple form can be seen as an interacting theory (due to the
appearance of a triple pomeron vertex~\cite{BW, Mueller, Bal, Kov, BV, BLV2}) of non local
Pomerons in 2+1 dimensions. Such field theory models, even if oversymplified,
are too complicated to be analyzed analytically.
What have been proposed in the past as even simpler toy models to be analyzed
and improve the understanding were models in 0+1 dimensions, a kind of quantum mechanics of
pomerons, and a local 2+1 dimensional theory. They were introduced before the
QCD analysis~\cite{schwimmer,acj}, but nevertheless are characterized by some common features
implied by it. 
The former was analyzed many years ago and reconsidered recently~\cite{KL,Bond,BVtoy} from
different points of view.
The latter was also studied many years ago but most of the questions remained open.
We report here some results of a recent work~\cite{BVtoy2} devoted to formulate a novel perturbative
approach useful to analyze this $2+1$ QFT model.  

\section{PT symmetry in QM and QFT}
It is well known that a partial effective description of a system can be
associated to a non hermitian Hamiltonian which is characterized by a non
unitary evolution. Nevertheless several attempts to analyzed such systems and to
try even to formulate some consistent non hermitian quantum mechanics has been done.
The first interesting result was found by Bender \cite{bender} who noted that there
exist non hermitian Hamiltonians having a real spectrum bounded from below.
This was shown to be possible if also specific boundary conditions for the
wave functions of the associated Sturm-Liouville differential problem were
properly defined.
The main point at the base of these properties is that such Hamiltonians can be
formally obtained by a similarity transformation acting on well defined
hermitian ones~\cite{mostafazadeh}. Note that the same considerations formally
apply to systems with a finite (QM) or infinite (QFT)~\cite{BenderQFT} degrees of freedom. 

A special class of Hamiltonians having these property was found to have an
unbroken $PT$ symmetry, so that the Hamiltonian $H$ and the operator $PT$ have
common eigenstates. In such a case it is possible to define a special scalar
product, leading to a norm conserved in time, as follows
$(f,g)=\int dx [PT f]^t(x) g(x)$, $t$ denoting the transpose operation.
Such $PT$-norm is not positive and there exist an operator $C$, commuting with
both $H$ and $PT$, which select the two possible signs.
By constructing the operator $C$, which depends on the Hamiltonian $H$ itself,
it is therefore possible to construct another scalar product
$\langle f|g \rangle=\int_\Gamma dx [CPT f]^t(x) g(x)$ and define an Hilbert
space with a positive norm conserved in time. The observables are defined to
satisfy $O^t=CPT O CPT$, which reduces to the hermiticity condition for the
usual case $C=P$ in conventional QM.

In our case the local effective theory for interacting pomerons ($PT$
symmetric and with an Hamiltonian with real eigenvalues) is associated to
the evolution in rapidity and the scalar product is the conventional one with
the norm of pomeron states, which interact by the triple pomeron vertex,
not conserved in rapidity. Nevertheless the
operators introduced above are very useful to develope a perturbative analysis. 

Indeed let us consider a system with the Hamiltonian
\beq 
H=H_0+\lambda H_I \,,
\eeq 
where $H_0$ (the free part) is hermitian and $H_I$ (the interaction part) is anti-hermitian.
We define the parity operator to transform $H$ into $H^\dagger$ so that
$[H_0,P]=0$,  $\{H_I,P\}=0$ and $P^2=1$. One has to look for the operator $C$
such that
\beq
[C,H]=[C,PT]=0\,.
\label{chpt}
\eeq
It is convenient to assume the general form $C=e^Q P$, where $Q$ is an
hermitian operator, that together with the previous commutation relations imply
\beq
2\lambda \, e^Q H_I=[e^Q,H]\,.
\label{eqQ}
\eeq 
Moreover one obtains easily the relation
$e^{-Q} H e^Q=H^\dagger$, 
which also implies
\beq
 h=e^{-Q/2} He^{Q/2}=e^{Q/2} H^\dagger e^{-Q/2}=h^\dagger\,.
\label{defh}
\eeq
Therefore we have found an hermitian Hamiltonian $h$ which is similar to $H$
by means of the similarity transformation induced by the operator $e^{Q/2}$. 

This general relations can be studied perturbatively for a small coupling
$\lambda$.
We start by looking for a perturbative expansion of
$Q=\lambda Q_1+\lambda^3 Q_3+...$ by solving eq. (\ref{eqQ}) which gives:
\beq
 [H_0,Q_1]=-2H_I,\ \
[H_0,Q_3]=-\frac{1}{6}\Big[[H_I,Q_1]Q_1\Big] \label{condq}
\eeq
and so on. From these relations one obtains, as we shall see,
an explicit form for the $Q_i$.
Once $Q$ is known as a power series in $\lambda$, the Hamiltonian $h$
can also be found in the same form:
$h=h^{(0)}+\lambda^2h^{(2)}+\lambda^4h^{(4)}+...$ with the first terms given by
\beq h^{(0)}=H_0, \quad
h^{(2)}=\frac{1}{4}[H_I,Q_1], \quad
h^{(4)}=\frac{1}{4}[H_I,Q_3]+\frac{1}{32}\Big[[H_0,Q_3]Q_1\Big]\,.
\label{hamilh}
\eeq

\section{Analysis of the LRFT model}
Let us start by defining the LRFT as a theory of two fields $\phi(y,x)$ and $\phid(y,x)$
depending on rapidity $y$ and transverse coordinates $x$ with a 
Lagrangian density
\beq
{\cal L}=\phid(\partial_y-\mu-\alpha'\nabla_x^2)\phi+i\lambda
\phid(x)\Bigl[\phid(x)+\phi(x)\Bigr]\phi(x),
\eeq
where $\mu>0$ is the intercept minus unity and $\alpha'$ is the slope
of the pomeron trajectory. Note that if $\mu>0$ the corresponding functional integral
is divergent and the only way to define the theory beyond the set of
perturbative Feynman diagram is the analytic continuation from $\mu<0$ when
the theory is well defined. Such a continuation is automatic in the the Hamiltonian approach,
where a quasi-Schroedinger equation for the wave function $\Psi$ is defined:
\beq
\frac{ d\Psi(y)}{dy}=-H\Psi(y), \quad H=H_0+\lambda H_I
\eeq
with the free part given by
\beq
 H_0=\int d^2x
(-\mu\phid(x)\phi(x)+\alpha'\nabla\phid(x)\nabla\phi(x))
\eeq 
and the interaction part by 
\beq 
H_I=i\int
d^2x\, \phid(x)\Bigl[\phid(x)+\phi(x)\Bigr]\phi(x)\,.
\eeq
Standard commutation relations are valid between $\phi$ and $\phid$:
$[\phi(x),\phid(x')]=\delta^2(x-x')$.
The scattering amplitude with the target ('initial') state $\Psi_i(y_1)$ at rapidity 
$y_1$
and the projectile ( 'final') state  $\Psi_f(y_2)$ at rapidity $y_2>y_1$ is defined as
\beq
iA_{fi}(y_2-y_1)=\langle \Psi_f(y_2)|e^{-H(y_2-y_1)}|\Psi_i(y_1)\rangle.
\label{ampl}
\eeq
One can demonstrate that the perturbation expansion in powers of $\lambda$ of this expression
reproduces the standard Reggeon diagrams of the LRFT and also that (\ref{ampl})
satisfies the requirement of symmetry between the target and projectile
(see ~\cite{BV})
Parity transformation $P$ is defined by
$\phi(y,x)\to -\phi(y,-x)$ and $\phid(y,x)\to -\phid(y,-x)$ so that $P H P=
H^\dagger$, while $T$ is the complex conjugation, so that $[H,PT]=0$.
Any state can be written as $F(i \phid)|0\rangle$, where $F$ is a real function and
$\phi |0\rangle=0$ so that $\langle\Psi|H|\Psi\rangle=\langle\Psi_0|F(-i\phi) H F(i \phid)|\psi_0\rangle$.
We shall look for a perturbatively constructed similarity transformation, as
illustrated in the previous section, in order to write any transition
amplitude as
\beq
iA_{fi}(y_2-y_1)=\langle e^{Q/2}\Psi_f(y_2)|e^{-h(y_2-y_1)}|e^{-Q/2}\Psi_i(y_1)\rangle.
\eeq
 In particular the simplest object one can imagine is the full pomeron
Green function at rapidity $y$ and momentum $k$ will be given as as
\beq
\delta^2(k-k')G(y,k)=<0|\phi(k)e^{Q/2}e^{-yh}e^{-Q/2}\phid(k')|0>.
\eeq
On restricting to the first non trivial order in perturbation theory one is
looking for
\beq
Q_1=-2\frac{i}{\mu}\int d^2x_1d^2x_2d^2x_3
\Big(f_1(x_1,x_2,x_3)\phid_1\phi_2\phi_3
- h.c.\Big)\,,
\eeq
and obtains for the Fourier transform of $f_1$:
\beq
\tilde{f}_1(k_1,k_2,k_3)=\mu\frac{(2\pi)^2\delta(k_1+k_2+k_3)}
{\mu-\alpha'(k_2^2+k_3^2-k_1^2)}\,.
\label{f1}
\eeq
The corresponding correction of order $\lambda^2$ to the hamiltonian $h$
similar to $H$ is given by
\beq
h^{(2)}=\frac{1}{4}[H_I,Q_1]=h^{(2)}_{single}+h^{(2)}_{pair}+h^{(2)}_{NC}\,.
\label{h2}
\eeq

The first contribution is a correction to the single pomeron propagation.
\beq
h^{(2)}_{single}=\int d^2k\phid(k)\phi(k)\Delta^{(2)}\epsilon(k)\, , \quad 
\Delta^{(2)}\epsilon(k)=-
\frac{2}{(2\pi)^2} {\rm Re}\,
\int \frac{d^2k_2d^2k_3\delta^2(k_2+k_3-k)}
{\mu-\alpha'(k_2^2+k_3^2-k^2)}\,.
\eeq
This term gives a correction to the pomeron energy
$\epsilon(k)=-\mu+\alpha'k^2+\lambda^2\Delta^{(2)}\epsilon(k)$.
A renormalization is needed and choosing the condition $\epsilon(0)=-\mu$ one
finds $\Delta^{(2)}\epsilon_{reg}(k)=-\frac{1}{8\pi \mu}k^2$ which leads to
the renormalization of the pomeron slope
\beq
\alpha'\to\alpha'_{ren}=\alpha'-\lambda^2\frac{1}{8\pi \mu}\,.
\eeq

The last term in eq. (\ref{h2}) $h^{(2)}_{NC}$ is not conserving the pomeron number and
therefore contributes to order $\lambda^4$ to the eigenvalues, going beyond
our approximations. We neglect it here.

The second term of eq. (\ref{h2}) $h^{(2)}_{pair}$ has a complicated structure
associated to the interaction of two pomerons
\beq
h^{(2)}_{pair}=\int d^2k_1d^2k_2d^2q_1d^2q_2\delta^2(q_1+q_2-k_1-k_2)
V^{(2)}(q_1,q_2|k_1,k_2)\phid(q_1)\phid(q_2)\phi(k_1)\phi(k_2)\,,
\eeq
with an interaction potential $V^{(2)}$ being non local and with some
degenerate terms depending only on the incoming or outgoing momenta.
In such a case one may be interested in studying the scattering states, not changing
the spectrum, and in the presence of bound states which instead could deeply
affect the spectrum.
Let us note that due to the fact that to order $\lambda^2$ in the spectrum of
$h$ there are no transition in the number of pomeron states, one can really
solve the problem with quantum mechanical techniques.
In the analysis of the two pomeron potential we have considered for simplicity
the forward direction $q_1+q_2=k_1+k_2=0$ so that
$V^{(2)}(q_1,q_2|k_1,k_2)=V(q,k)=v(q)+v(k)+V_1(q,k)$ with
$v(q)=\frac{1}{8\pi^2}\frac{1}{\mu-2\alpha'_{ren} q^2}$ and
$V_1(q,k)=-\frac{1}{2\pi^2}\frac{1}{\mu-2\alpha'_{ren}
  (k^2+(k+q)^2-q^2)} + (k \leftrightarrow q)$.
Therefore the
Schr\"odinger equation to be solved reads in momentum space reads
\beq
(\epsilon(q)-E)\psi(q)=-\int d^2kV(q|k)\psi(k)\,.
\label{eq1}
\eeq
Omitting the technical details derived in \cite{}, we simply present here our findings.
Solving the associated Lippman-Schwinger equation for the scattering matrix
one obtains to order $\lambda^2$, after performing a regularization to handle divergent quantities, 
\beq
T(q|l)=V_1(q|l)-\frac{v(l)}{I_2}\chi_2(q)-\frac{v(q)v(l)}{I_2}\,,
\eeq
where
$\epsilon(q)=-2\mu+2\alpha'_{ren} q^2$, $I_n=\int d^2k
\frac{v^n(k)}{\epsilon(l)-\epsilon(k)}$
and $\chi_2(q)=\int d^2k\frac{V_1(q,k)v(k)}{\epsilon(l)-\epsilon(k)}$.
This also gives the solution of the scattering states
$\psi_l(q)=\delta^2(q-l)+\frac{T(q|l)}{\epsilon(l)-\epsilon(q)\pm i0}$.

In order to investigate the existence of bound states of energy $E$ we consider the
associated equation
\beq
t_E(q)=\int d^2k\frac{V(q|k)t_E(k)}{E-\epsilon(k)}\,, \quad \psi_E(q)=\frac{t_E(q)}{E-\epsilon(q)}\,.
\label{eq4}
\eeq
The condition of the existence of bound states can be reduced to the existence
of the solution of a secular equation of a finite algebraic problem.
in a perturbative sense, i.e. for small values of $\lambda$ one can show that
there are no solutions. They may appear at larger values of $\lambda$, a case
for which  which nevertheless some higher order terms in the perturbative expansion
may be also important.
We stress that the results obtained are valid for an evolution along rapidity
intervals $\Delta y \sim 1/\lambda^2$.
%

\begin{footnotesize}

\end{footnotesize}
\end{document}